# Time-resolved diffusing wave spectroscopy applied to dynamic heterogeneity imaging

M. Cheikh, H.L. Nghiêm, D. Ettori, E. Tinet, S. Avrillier and J.-M. Tualle*

*Laboratoire de Physique des Lasers, Centre National de la Recherche Scientifique, Unité Mixte de Recherche (CNRS UMR 7538),*

*Université Paris 13, 99 avenue J-B. Clément, 93430 Villetaneuse, France*

**Abstract**

We report in this paper what is to our knowledge the first observation of a time-resolved diffusing wave spectroscopy signal recorded by transillumination through a thick turbid medium: the DWS signal is measured for a fixed photon transit time, which opens the possibility of improving the spatial resolution. This technique could find biomedical applications, especially in mammography.

120.6160  Speckle interferometry
170.3340  Laser Doppler velocimetry
170.3880  Medical and biological imaging

OCIS codes: 120.6160, 170.3340, 170.3880

* Corresponding author email address: tualle@galilee.univ-paris13.fr

Near infrared spectroscopy (NIRS) is of special interest for breast cancers screening. This inoffensive technique, using non-ionizing radiation, is indeed very sensitive to the presence of hemoglobin in the tissue and has been shown to allow tumors detection[1-5] with a detection efficiency greater than 90% when time-resolved measurements are performed in the transmittance geometry[4,5]. The use of time-resolved measurements definitely presents several advantages: the spatial resolution is better at short times, while later times, which are less sensitive to edge effects, allow a better discrimination of the absorption coefficient. However, the main limitation of this technique remains its specificity since the number of false-positives cases is still too high for screening[5]. The use of other contrast factors should therefore considerably improve these clinical results.

Diffusing Wave Spectroscopy (DWS) which is highly sensitive to the microscopic movements in the medium could provide such a contrast factor. The possibility of imaging a dynamic heterogeneity with DWS was previously reported[6,7] with typical scales up to 20 transport mean free paths. Provided that a signal could be registered in transillumination through a thickness that is close to 40 transport mean free path, such a technique could make available an interesting complementary information about capillaries blood flow in the field of mammography.

We report in this paper the observation of a time-resolved diffusing wave spectroscopy signal recorded by transillumination through a thick turbid medium. For these measurements, we used an interferometric method that we recently developed [8-10] to perform diffuse light time-resolved measurements . This technique, which opens the possibility of recording DWS signals for fixed photon path-lengths up to 300 mean free paths,[11] therefore combines the advantages of time-resolved diffuse intensity measurements and time-resolved DWS.

Time resolved measurements are performed[8-10] using an interferometer (fig 1) and a wavelength modulated continuous laser source. The sinusoidal modulation at frequency $f$, with $\omega(t) = \omega_0 + \Delta\Omega \cos(2\pi f t)$, allows to simulate an incoherent source. The recorded fluctuations $s(t)$ of the speckle pattern (after rejection of the DC components) are analyzed in the following way: we compute

$$S_{DC,m}(\tau) = 2f \int_{m\Delta T/2}^{(m+1)\Delta T/2} s(t) \mathrm{Ref}(t,\tau) \, dt, \qquad (1)$$

where $m$ is an integer, $\Delta T = f^{-1}$ is a modulation period, and where

$$\text{Ref}(t,\tau) = \sin^4(2\pi f t) \exp[i \Delta\Omega \tau \cos(2\pi f t)] \qquad (2)$$

For DWS the main result [11] concerns the following ensemble average:

$$I_p(\tau) = < S_{DC,m}(\tau) S^*_{DC,m+2p}(\tau) > \propto I(\tau) g_1(t = p\Delta T, \tau) \qquad (3)$$

where $I(\tau)$ is the time-resolved diffuse intensity (transmitted or back scattered), and $g_1$ is the normalized field correlation function. For $p=0$, $g_1 = 1$ and $I_0(\tau) = I(\tau)$.

Let us point out that two time-scales are present in this function: $t$ represents the correlation time, in the millisecond range, while $\tau$ is the photon time of flight, in the hundred picoseconds range. The correlation function exhibits an exponential behavior[12-14] with $\tau$:

$$g_1(t, \tau) = \exp[-\mu_f(t) c \tau] \qquad (4)$$

where $c$ is the light speed in the medium, and where $\mu_f$ is an effective absorption coefficient which depends on the microscopic movements of the scatterers and on time $t$. DWS therefore allows a new kind of contrast, that can be varied only by varying time $t$.

The experimental set-up, detailed in previous papers[8-10], is schematically shown in fig 1. The light source is a 7 mW wavelength modulated laser diode emitting at 780 nm, with a line width of about 1 MHz. This source allows for $f=300$ Hz a modehop-free modulation of 6 Ghz, which corresponds to a time resolution of 230 ps. The whole experiment was performed using graded-index multimode fibers for both the signal and the reference arms. The dispersion introduced by these fibers is negligible compared with the time resolution. The signal-to-noise ratio is improved by two ways: The reference beam fluctuations are reduced by the use of an optical isolator, a low reference signal, and a balanced detection; In order to cancel parasitic signals, a periodic background subtraction is performed using an acousto-optic modulator inserted in the measurement arm.

The sample is a breast-like phantom made of a suspension of calibrated polystyrene microspheres (diameter: 520 ± 37 nm, refractive index: 1.580 at 780 nm) in glycerol. Such a viscous liquid prevents field decorrelation during the modulation period $f$.[11] The reduced scattering coefficient was adjusted to be around *9 cm$^{-1}$* since most published data[15-17] for breast tissues scattering coefficient range from *6* to *11 cm$^{-1}$* near 800 nm. The corresponding values for the absorption coefficient vary[15-17] from *0.02* to *0.08* cm$^{-1}$, so that glycerol, which has a low absorption coefficient in the near infrared[18] ($\mu_a$ = *0.02 cm$^{-1}$* at 780 nm), is a good candidate for simulating the bulk. The optical coefficients of the sample were counterchecked with time-resolved transmittance measurements: a fit with a diffusion model led to $\mu_s'$=9.5±0.2 cm$^{-1}$ and $\mu_a$=0.026±0.001 cm$^{-1}$, in good agreement with the expected value.

The liquid phantom fills an optical glass container (*4×6×9.5 cm*). A thin cell (*0.1×1×4 cm*) is inserted in this background medium in order to simulate an heterogeneity. This cell is filled by a suspension with the same optical properties as the background, using the same calibrated microspheres in water. The point is that water is much less viscous than glycerol, and will induce a higher decorrelation due to Brownian motion: we therefore await from correlation measurements a drastic increase of the effective absorption coefficient associated to this dynamical heterogeneity.

Time resolved measurements were performed in the transmittance geometry, as shown in figure 2: the beam is perpendicular to the illuminated surface of the sample. The emission and detection fibers face each other, separated by the L = 4 cm thickness of the container. The cell is placed in the middle of the container, half–way between the source and the detector. Its smallest side is parallel to the source-detector axis, the largest side being vertical, so that the heterogeneity can be considered as infinite in this direction. It can move in the horizontal direction orthogonal to the source-detector axis. Such a translation is somewhat equivalent to a translation of the source-detector system, and was chosen for practical convenience.

The speckle pattern fluctuations *s(t)* of the transmitted light were registered for different positions *x* of the cell center along the horizontal axis, ranging from *0* to *3 cm* with *3 mm* steps. The measurements of the transmitted intensity $I_0(\tau) = I(\tau)$ and of $I_2(\tau)$ (*p=2* in equation 3) were performed through numerical processing with an ensemble averaging over 90 000 modulation periods. The ratio of the time-resolved curves recorded with *p=0* and *p=2* leads directly to $g_1(t, \tau)$ at position *x* and with $t = 2/f = 6.67\,ms$.

Figure 3 presents the results obtained for $I$ and $g_1$ versus $x$, at $t = 6.67\, ms$ and for $\tau = 756\, ps$. The left axis gives the scale for $g_1$, while the right axis corresponds to $I$. It clearly appears that the object cannot be seen from transmittance measurements, which is not surprising since the optical properties of the cell and the background are identical. However a contrasted shadow of the object can be seen in the correlation function profile due to the different mobility of the microspheres in water. The correlation curve $g_1$ can be fitted by a Gaussian, with a full width at half maximum (FWHM) $\Delta x = 1.51\, cm$.

The second point we want to experimentally show here is that the spatial resolution is improved at short times $\tau$. Let us clearly define this quantity: in the framework of the first order Born approximation, the perturbation of the transmitted signal induced by the heterogeneity corresponds[19] to the convolution product, along the $x$ axis, of the profile $\Pi$ of the object and a kernel $K_\tau$. The spatial resolution $R(\tau)$ is the FWHM of this kernel. In our case $\Pi$ is rectangular, with a width $l = 1cm$, while $K_\tau$ is almost Gaussian. The convolution product $\Pi \otimes K_\tau$ can then be explicitly written using the error function[20] and it can be easily shown that, for high values of $R(\tau)$, a good approximation of $\Pi \otimes K_\tau$ can be obtained by using for $\Pi$ a Gaussian profile with a FWHM :

$$W = 2\sqrt{\frac{\ln 2}{6}}\ l \approx 0{,}68 l \qquad (5)$$

This approximation is quite good in the whole range of the experiment, so that the experimental resolution $R(\tau)$ can be calculated from the measured values of $\Delta x(\tau)$ using:

$$R(\tau)^2 = \Delta x(\tau)^2 - W^2 \qquad (6)$$

$R(\tau)$ is presented on figure 4 as a function of $\tau$, together with its theoretical values obtained using[21]:

$$R_{th}(\tau, z = L/2) = \frac{2.35}{\sqrt{2}}\sqrt{\frac{c\Delta t}{3\mu_s'}} \qquad (7)$$

We find a good agreement between theory and experiment, that is to say a clear improvement of the spatial resolution at short times. We notably have a space resolution of about 1 cm for $\tau = 400$ ps, to be compared with a space resolution greater than 1.7 cm in the CW case. This is a quite realistic demonstration of the potentialities of this technique, which combines the advantages of time resolved measurements and DWS.

**Conclusion**

We report what is to our knowledge the first observation of a time-resolved DWS signal in transillumination through a thick medium. An improvement of the spatial resolution was observed at shorter times. This technique could find an application for mammography, with a new kind of contrast factor linked to the capillaries blood flow . An improvement of the setup should now consist in multiplying the number of detectors, using for instance a CCD camera[8], in order to obtain a better sensitivity.

# FIGURE CAPTIONS:

Figure 1:

Experimental set-up. The source is a wavelength modulated laser diode, multimode fibers are used for both the signal and the reference arms; beam fluctuations are reduced via an optical isolator and a balanced detection. An acousto-optic modulator (AOM) allows a periodic background subtraction.

Figure 2:

The sample is an optical glass container of thickness L=4 cm filled with a breast-like liquid phantom; a thin cell (*0.1×1×4 cm*), filled by a suspension with the same optical properties, is inserted in the middle of the container and can move in the horizontal direction (x axis) orthogonal to the source-detector z axis.

Figure 3:

Measured transmitted intensity I(x) (right axis and solid triangles) and correlation function $g_1(x)$ (left axis and solid circles) for a fixed photons transit time $\tau$ = 756 ps. The Gaussian fit of $g_1(x)$ is indicated by a dashed line.

Figure 4:

Experimental resolution $R(\tau)$ as a function of the photons transit time $\tau$, together with its theoretical values (dashed line).

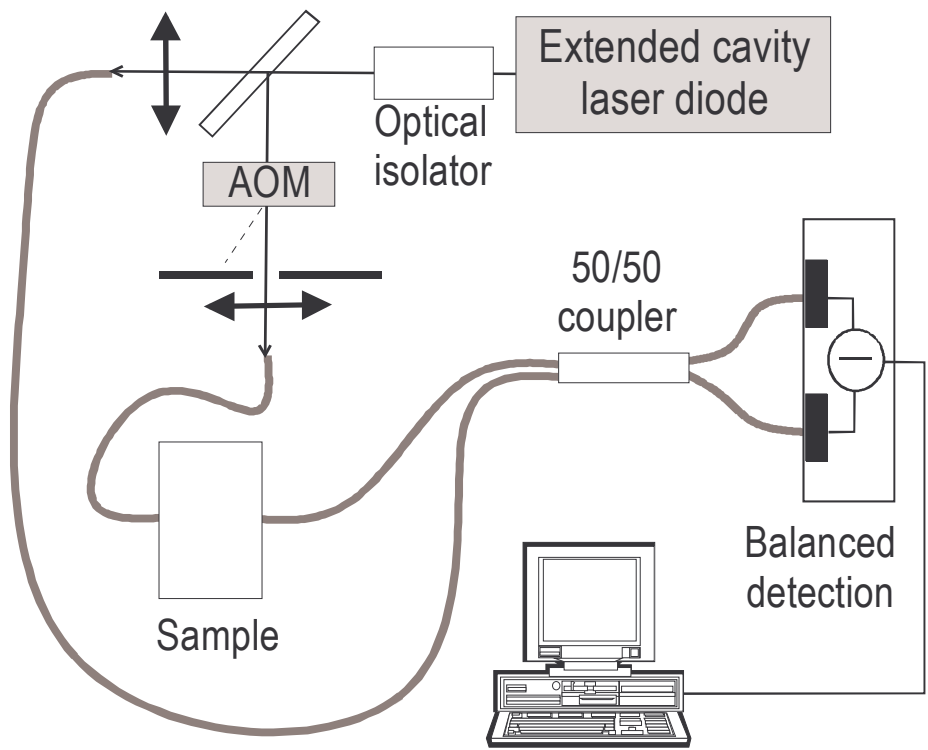

Figure 1

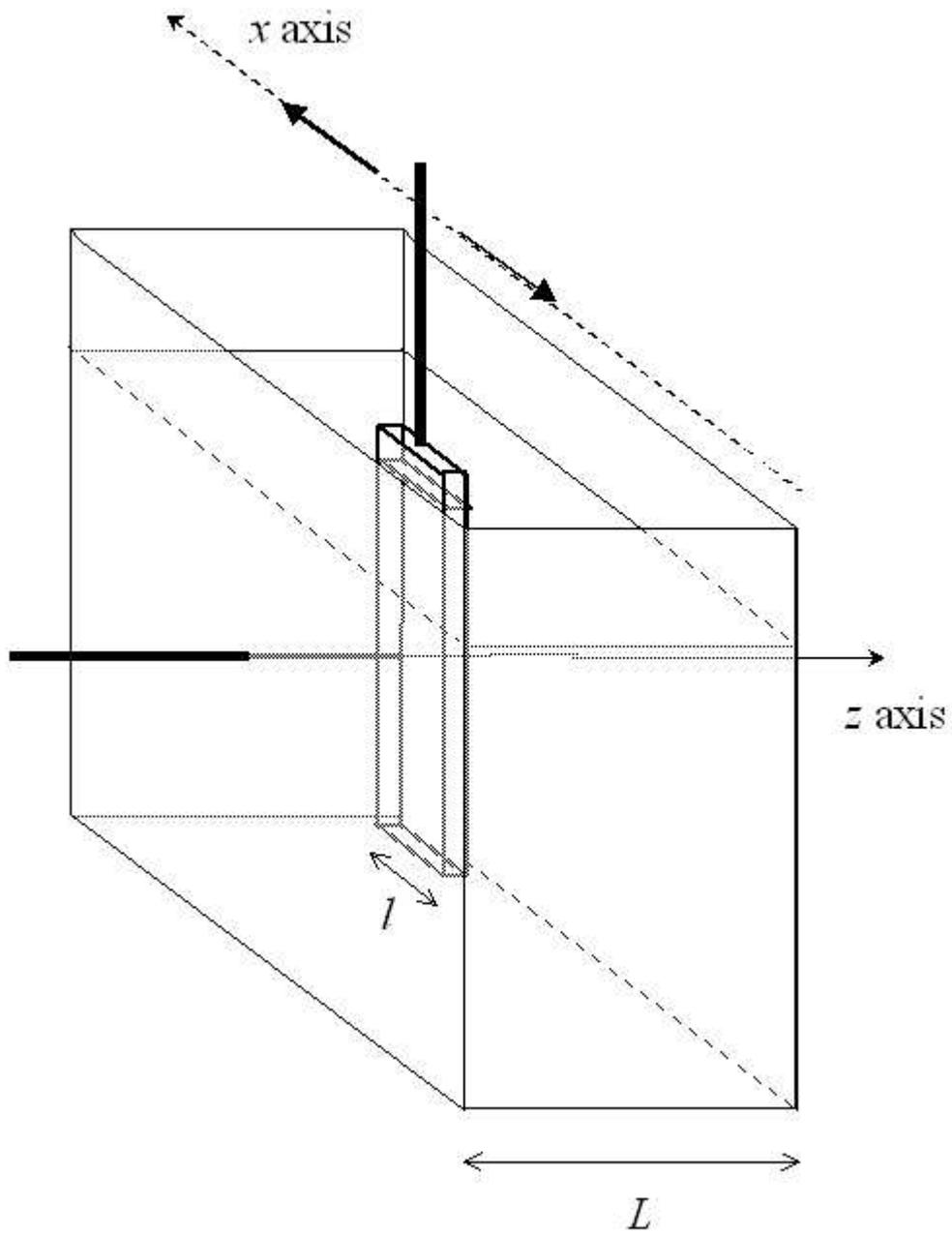

figure 2

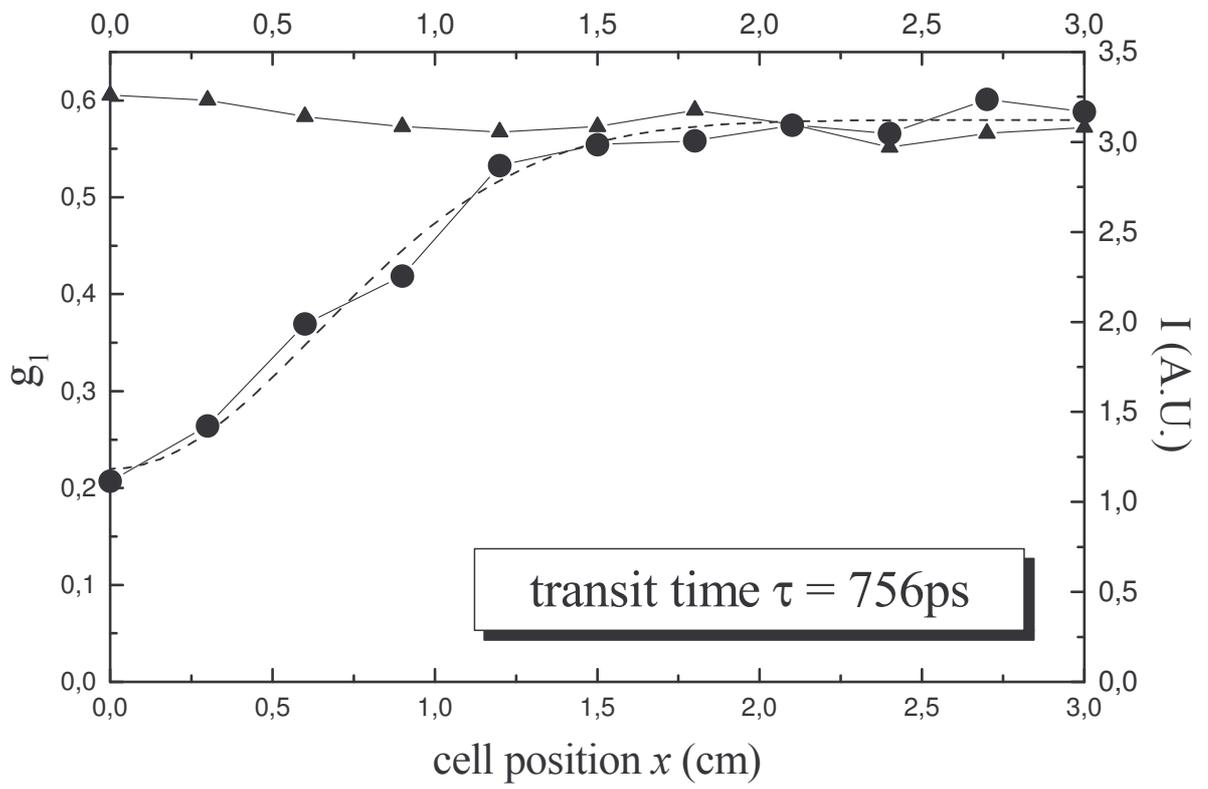

figure 3

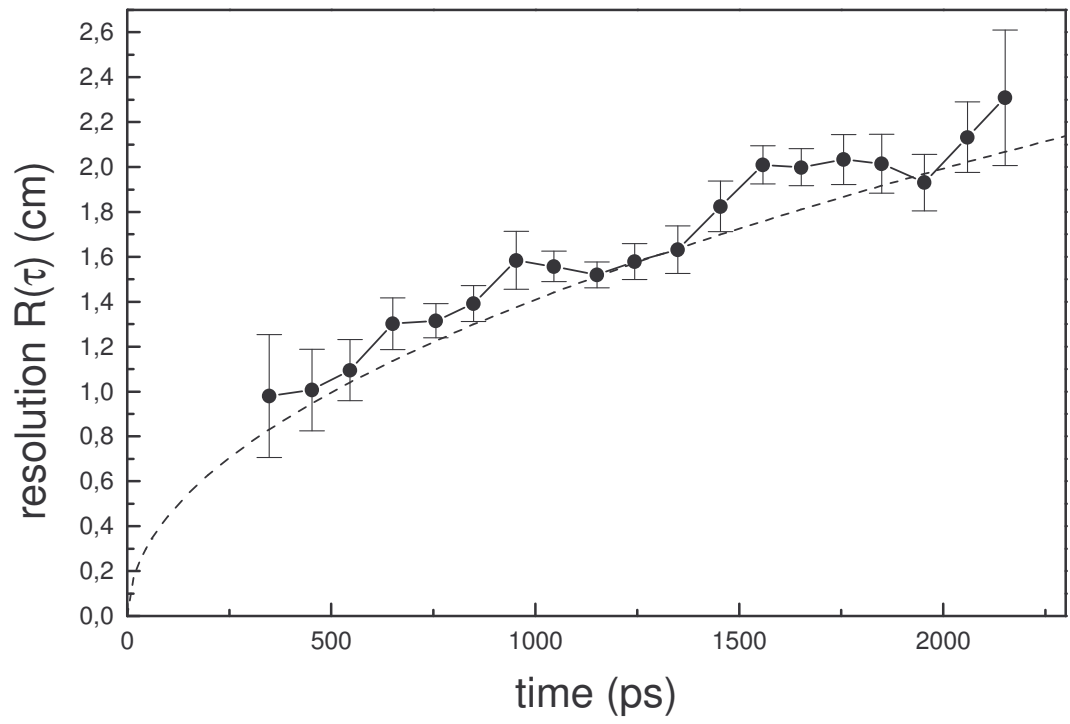

Figure 4